\documentclass[10pt, conference, draftclsnofoot, onecolumn]{IEEEtran}
\usepackage{cite}
\usepackage{longtable, ltcaption}
\usepackage{multicol}
\usepackage{comment}
\usepackage{algorithm,algpseudocode}

\usepackage{mathtools}
\usepackage{subfigure}
\usepackage{color}
\usepackage{float}
\usepackage{multirow}
\usepackage{amsmath}
\usepackage[justification=centering]{caption}
\usepackage{caption}

\ifCLASSINFOpdf
   \usepackage[pdftex]{graphicx}
\else
\fi
\IEEEoverridecommandlockouts

\begin{document}
%
\title{Identifying a Criminal's Network of Trust}


\author{\IEEEauthorblockN{Pritheega~Magalingam, Asha~Rao~\IEEEmembership{Member,~IEEE}, Stephen~Davis}
\IEEEauthorblockA{School of Mathematical and Geospatial Sciences\\
RMIT University, Melbourne, Australia\\
GPO Box 2476, Melbourne, Victoria 3001\\
Email: pritheega.magalingam@rmit.edu.au, asha@rmit.edu.au, stephen.davis@rmit.edu.au\\
}
}


%


\maketitle

\begin{abstract}
Tracing criminal ties and mining evidence from a large network to begin a crime case analysis has been difficult for criminal investigators due to large numbers of nodes and their complex relationships. In this paper, trust networks using blind carbon copy (BCC) emails were formed. We show that our new shortest paths network search algorithm combining shortest paths and network centrality measures can isolate and identify criminals' connections within a trust network. A group of BCC emails out of 1,887,305 Enron email transactions were isolated for this purpose. The algorithm uses two central nodes, most influential and middle man, to extract a shortest paths trust network.
\end{abstract}


%
\IEEEpeerreviewmaketitle

\section{Introduction}
The task of identifying links and location of a criminal within a network of a large set of people is a difficult one \cite{sarvari2014}. Often it is easier to identify criminals by taking a subset of the given network \cite{yasin2014}. In this paper, particular subsets of the emails from the Enron email database, namely emails with bcc recipients are used to isolate possible criminal trust subnetworks. We combine social network measures with the shortest path algorithm to achieve this aim. 

Social network analysis is regularly used to investigate and view communities as networks of individual relationships that consist of people or organisations that are grouped using their common features \cite{newman2010}. Such networks represent entities as nodes or vertices, depicting the communication or other relationships among them via edges.

A number of tools are used to analyse networks. The shortest path algorithm \cite{Cormen2009} is a graph theoretic tool used in social network analysis to retrieve information about individual relationships and to identify the strongest connection between nodes in a criminal network \cite{xu2004, yunkai2006}. Strong relationship between nodes depends on the different purposes and roles of each node. A central node is defined as the most important node based on various factors, for example, on the number of in-coming or out-going connections, the number of times the node appears in all paths, the number of times the node appears adjacent to a target node, etc. See, for example \cite{freeman1991}. 

In this paper, we use the Enron database which contains more than a million email transactions. Many researchers from different backgrounds have worked on various aspects of the Enron data to detect important nodes\cite{yang2010}, and business processes \cite{shetty2005}, identify manager-subordinate relationships \cite{diehl2007} and explore organisational behaviour \cite{diesner2005}. In \cite{gao2012}, the total weight of shortest paths of Enron email hypergraphs was used to identify important persons. Shetty et al. \cite{shetty2005} predict the occurrence of an email sequence of a given length and show that choosing different lengths of sequences gives different sets of influential nodes within the group of higher authorities. Email transaction shortest paths between Enron employees have also been studied by various researchers. Tang et al. \cite{tang2010} analyse the shortest path according to the time taken by a sender to deliver an email to a recipient using spread of a message to show the importance of a person to retain a communication. Similarly Yelupula et al. \cite{yelupula2008} used email flow and numerical analysis to predict the hierarchy of Enron management. One of the features used was the number of email addresses in the TO, CC and BCC fields. Hershkop \cite{hershkop2006} suggests that if it was possible to find a rule to represent the appearance of To, CC and BCC addresses then it would be easy to distinguish the pattern which violates the norms. Stolfo et al. \cite{stolfo2006} compile a list of recipients of a particular sender and reduce it by removing all the duplicates. This list was later divided into different subsets which were used to model user cliques. Any new clique that violates the normal pattern is detected as abnormal. List of recipients' accounts have also been used to determine communication behaviour measures \cite{shuklaanalysis}. 

Distinct from other research, our paper uses only emails that have at least one BCC recipient. The main reason for choosing BCC recipients is because the `to' and `cc' recipients are visible to every recipient, while `bcc' recipients are inherently secretive \cite{uscertgovonBCC, bogawar2012}. Keeping the recipients concealed and not revealed to the other recipients creates a trust between the sender and BCC recipients \cite{fox2012}. This trust can be based on certain types of relationships, for example a bcc recipient can be a friend, a business partner, manager, etc. to the sender \cite{fox2012}. Categorizing trust level involves different subject areas for example, a person's skill, knowledge, behaviour \cite{golbeck2003}, recommendation or opinion \cite{josang2008}. 

There are no studies on usage of only those emails that have BCC recipients to form a trust network. In this paper, we isolate groups of criminals who are linked in this trust network. For this purpose, we show that the shortest path concept together with centrality measures can be used for grouping and possibly identifying criminal connections in a network, using the publicly available Enron email dataset \cite{ISI:2009} and corroborating our work with published information about the crimes that led to the collapse of Enron \cite{Kathleen2003}.

In the next section, we give the background for our work, including network centrality measures, shortest paths and details of the crimes that occurred within Enron.  In section \ref{sec:EmailGroupingandThresholdLimit} we present our statistical analysis of Enron BCC recipients and further tighten the scope of our research. Section \ref{sec:AnalysisAlgorithm} gives the shortest path network search algorithm. Section \ref{ShortestPathsNetworkAnalysis} discusses the experiments conducted and the discovery of several shortest paths trust networks. In Section \ref{sec:firstsuspectTest}, we present a test to show that the algorithm works well in a different investigative scenario and in Section \ref{sec:DiscussionofResults}, we dicsuss the results obtained. Finally we give the conclusion and future work. 

\section{Background}
\label{sec:Background}

A network is a graph that consist of vertices represented as points and edges as lines that connect its end vertices \cite{newman2010}. Our analysis involves unique steps to obtain the criminal links within the BCC network. Many different centrality measures are used in social network analysis to identify nodes of importance (central nodes). One such important node is the node with highest eigenvector centrality, called the hub. A node has high \textbf{eigenvector centrality} when it is connected to many nodes that have high degree \cite{newman2010}. This measure is an indicator of the popularity that tends to identify centers of large cliques \cite{bonacich2007}. Hansler \cite{henseler2010} in his project calculates eigenvalues to rank people based on the number of emails sent and received in an email network. Kayes et al. \cite{kayes2012} use the eigenvector centrality measure to find influential bloggers in a blogging community. Influential bloggers are the nodes with useful information related to a topic that are also connected to other nodes of the same type.  

Another important node that we consider is the node with highest betweenness centrality. \textbf{Betweenness centrality} of a node is equal to the number of times the node appears in the shortest paths that bridge one node to another or one component to another \cite{newman2010}. Tayebi et al. \cite{tayebi2011} rank criminals using the betweenness centrality measure and identify a key player in a co-offending network.

Arun et al. \cite{maiya2010} propose an algorithm that greedily selects nodes in a variety of situations such as in a co-authorship network to achieve maximum sampling and combine  several different centrality measures including eigenvector centrality and betweenness centrality to rank influential nodes. In our research, we call the node with highest eigenvector centrality as the ``most influential" (MI) and take a node with high betweenness to be playing the role of a ‘middleman’ (MM). The shorter the paths from a criminal to the MI, the less isolated is the criminal, indicating that it is positioned in between highly linked nodes. The nearer a criminal to the MM, the higher the possibility the criminal could be using the MM to influence other nodes or criminals. Thus, we find the shortest paths from criminal to nodes with highest betweenness centrality and eigenvector centrality to form a closely connected network.

A network path consists of a sequence of nodes connected by edges \cite{newman2010}. For example \{${v_1,e_1,v_2}$\} represents a path between two nodes ${v_1}$ and ${v_2}$ with ${e_1}$ the edge between ${v_1}$ and ${v_2}$ \cite{newman2010}. Connecting multiple paths forms a network. A path in a network can be categorised as directed or undirected. A directed path is an edge that comprises of a start node which points to a specific end node. An undirected path consist of an edge connecting two nodes with no direction pointing either way \cite{newman2010}. In this research, we will focus on directed paths. The length of a path is the number of hops from the start node to the end node. A node, ${v_i}$ is located one hop away from node ${v_j}$, if node ${v_j}$ is adjacent to node ${v_i}$, that is there an edge from ${v_i}$ to ${v_j}$. Thus, the path length from node ${v_i}$ to ${v_j}$ is 1 \cite{newman2010}. Similarly, path ${v_i}$ to ${v_j}$ has length three if the number of edges or hops from ${v_i}$ to ${v_j}$ is 3.

A network of $n$ nodes and $m$ edges may contain multiple paths connecting one node to another. If multiple paths exist from ${v_i}$ to ${v_j}$, a shortest path is a geodesic path between these two nodes such that no shorter path exists \cite{newman2010}. The terms detailed above are used throughout this paper.

\subsection*{History of Enron and its Collapse}
\label{sec:AndrewFastowAllegations}

In July 1985, Enron Corporation was created by Kenneth Lay through merging Houston Natural Gas, a utility company, and Omaha-based InterNorth, a gas pipeline company. Kenneth Lay became the Chief Executive Officer (CEO) and Chairman of Enron \cite{salter2008innovation}. From 1988 to 2001, Enron opened branches in many countries, expanding its business and becoming the middleman for energy trading between the United Kingdom, Europe, South America and India \cite{salter2008innovation}. Enron's hidden aim \cite{salter2008innovation} was to earn commissions through regulating the fluctuating energy price in the market, which was due to increasing competition between old and new suppliers. In early 1990, Enron's Gas Bank helped Enron Corporation emerge from a business of piping gas to secure a prominent place in energy tradings between suppliers and consumers \cite{salter2008innovation}. In the 1990s, Enron became an exceptionally large player in the United States' energy market. In 1998, Andrew Fastow was employed as the Chief Financial Officer of Enron \cite{salter2008innovation}.

After the colapse of Enron in 2001, Andrew Fastow was convicted of planning and designing a complex web of offshore partnerships and questionable accounting practices \cite{salter2008innovation}. U.S authorities seized up to USD 23 million United States' assets from Andrew Fastow and his family members. The report by Thomsen and Clark from Division of Enforcement, U.S. Securities and Exchange Commission \cite{secshamtransaction} shows that the cash came from illegal business transactions that Andrew Fastow arranged \cite{secshamtransaction}. Andrew Fastow was shown to be involved in covert side deals, earning money through sham transactions or money laundering \cite{secshamtransaction}. Besides Andrew Fastow, 9 others \cite{Kathleen2003} (Table \ref{table:MoneyLaunderingCriminals}) were convicted of money laundering. Each had a different involvement in the money laundering crime. In this paper we extract the ego networks of these criminals and construct a new shortest path network, aiming to indicate a trusted sub-network.


\begin{table}[H]
\centering
\caption{Enron money laundering criminals}
\label{table:MoneyLaunderingCriminals}
\begin{IEEEeqnarraybox}[\IEEEeqnarraystrutmode\IEEEeqnarraystrutsizeadd{2pt}{1pt}]{v/c/v/r/v/c/v}
\IEEEeqnarrayrulerow\\
& \mbox{{\bf Name}} && \mbox{{\bf ID}} && \mbox{{\bf Email Address}} &\\
\IEEEeqnarraydblrulerow\\
\IEEEeqnarrayseprow[3pt]\\
& \mbox{Andrew Fastow} && \mbox{686} && andrew.fastow@enron.com &\IEEEeqnarraystrutsize{0pt}{0pt}\\
\IEEEeqnarrayseprow[3pt]\\
\IEEEeqnarrayrulerow\\
\IEEEeqnarrayseprow[3pt]\\
& \mbox{Andrew Fastow} && \mbox{687} && andrew.fastow@ljminvestments.com  &\IEEEeqnarraystrutsize{0pt}{0pt}\\
\IEEEeqnarrayseprow[3pt]\\
\IEEEeqnarrayrulerow\\
\IEEEeqnarrayseprow[3pt]\\
& \mbox{Lea Fastow} && \mbox{11010} && lfastow@pop.pdq.net  &\IEEEeqnarraystrutsize{0pt}{0pt}\\
\IEEEeqnarrayseprow[3pt]\\
\IEEEeqnarrayrulerow\\
\IEEEeqnarrayseprow[3pt]\\
& \mbox{Lea Fastow} && \mbox{11009} &&  lfastow@pdq.net  &\IEEEeqnarraystrutsize{0pt}{0pt}\\
\IEEEeqnarrayseprow[3pt]\\
\IEEEeqnarrayrulerow\\
\IEEEeqnarrayseprow[3pt]\\
& \mbox{Kevin Hannon} && \mbox{10068} && kevin.hannon@enron.com  &\IEEEeqnarraystrutsize{0pt}{0pt}\\
\IEEEeqnarrayseprow[3pt]\\
\IEEEeqnarrayrulerow\\
\IEEEeqnarrayseprow[3pt]\\
& \mbox{Kenneth Rice} && \mbox{9994} && kenneth.rice@enron.com  &\IEEEeqnarraystrutsize{0pt}{0pt}\\
\IEEEeqnarrayseprow[3pt]\\
\IEEEeqnarrayrulerow\\
\IEEEeqnarrayseprow[3pt]\\
& \mbox{Rex Shelby} && \mbox{15224} &&  rex.shelby@enron.com  &\IEEEeqnarraystrutsize{0pt}{0pt}\\
\IEEEeqnarrayseprow[3pt]\\
\IEEEeqnarrayrulerow\\
\IEEEeqnarrayseprow[3pt]\\
& \mbox{Rex Shelby } && \mbox{15225} &&  rex\_shelby@enron.net  &\IEEEeqnarraystrutsize{0pt}{0pt}\\
\IEEEeqnarrayseprow[3pt]\\
\IEEEeqnarrayrulerow\\
\IEEEeqnarrayseprow[3pt]\\
& \mbox{A.Khan} && \mbox{205} &&  adnankkhan@hotmail.com  &\IEEEeqnarraystrutsize{0pt}{0pt}\\
\IEEEeqnarrayseprow[3pt]\\
\IEEEeqnarrayrulerow\\
\IEEEeqnarrayseprow[3pt]\\
& \mbox{Michael Kopper} && \mbox{12708} &&  michael.kopper@enron.com  &\IEEEeqnarraystrutsize{0pt}{0pt}\\
\IEEEeqnarrayseprow[3pt]\\
\IEEEeqnarrayrulerow\\
\IEEEeqnarrayseprow[3pt]\\
& \mbox{Ben Glisan} && \mbox{1369} &&  ben.glisan@enron.com  &\IEEEeqnarraystrutsize{0pt}{0pt}\\
\IEEEeqnarrayseprow[3pt]\\
\IEEEeqnarrayrulerow\\
\IEEEeqnarrayseprow[3pt]\\
& \mbox{Joe Hirko} && \mbox{8716} &&  joe.hirko@enron.com  &\IEEEeqnarraystrutsize{0pt}{0pt}\\
\IEEEeqnarrayseprow[3pt]\\
\IEEEeqnarrayrulerow\\
\IEEEeqnarrayseprow[3pt]\\
& \mbox{S.Yaeger} && \mbox{861} && anne.yaeger@enron.com  &\IEEEeqnarraystrutsize{0pt}{0pt}\\
\IEEEeqnarrayseprow[3pt]\\
\IEEEeqnarrayrulerow
\end{IEEEeqnarraybox}
\vspace{0.7em}
\begin{footnotesize}
  \begin{minipage}{\linewidth}%
    \small Table \ref{table:MoneyLaunderingCriminals} shows the list of criminals involved in Enron money laundering crime \cite{Kathleen2003, ISI:2009}. The ID in the table is a computer generated number assigned to distinct email addresses.
  \end{minipage}%
\end{footnotesize}
\end{table}

Even though there are 10 criminals in this list, there are more than 10 email addresses as some of the criminals have more than one email address. There are distinct links from each of these email addresses to different recipients, hence we do not merge the email addresses of a particular criminal.

\vspace*{1mm}

\section{Email Grouping and Threshold Limit}
\label{sec:EmailGroupingandThresholdLimit}

\vspace*{0mm}

\begin{figure}[H]
\DeclareGraphicsExtensions{.pdf,.png,.jpg,.eps,.tif}
\centering
\includegraphics[width=2.9in]{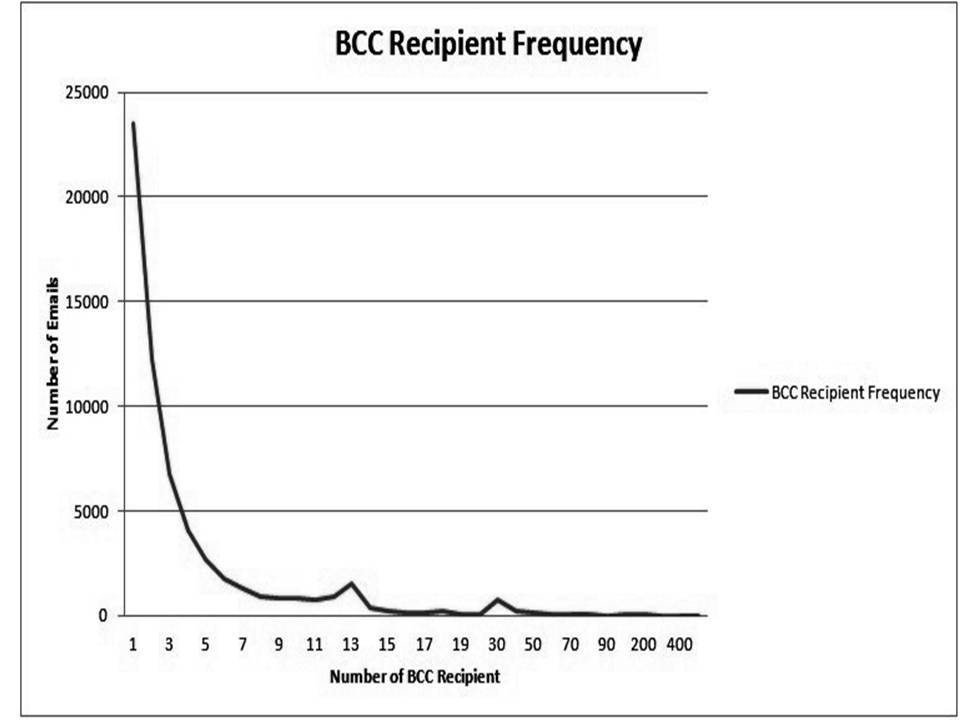}
\captionsetup{singlelinecheck=off,justification=justified}
\caption{Number of BCC recipients in emails. The majority of emails have at most 5, that is, 1 to 4 BCC recipients.}
\label{fig:NumberofBCCRecipientsinemails}
\end{figure}

Yasin et al. \cite{yasin2014} point out that taking only a part of a large amount of data from a network helps to reduce the complexity of identifying criminals. A subset  that occurs within the Enron email dataset are those that have BCC recipients. The bcc-ed email carries the email addresses of some recipients who are kept concealed from other recipients. We will use this categorisation to narrow the search scope, since within the 1,887,305 emails, there are 60649 emails with BCC recipients, giving a sizeable subset. We start with statistical analysis to further reduce our search space. Within the bcc-ed emails, a large group, 17784 emails, had 33.73\% of recipients bcc-ed ; that is, approximately 1 out of 3 recipients were bcc-ed. From Figure \ref{fig:NumberofBCCRecipientsinemails}, it is clear that the majority of emails had at most 5; 1 to 4 BCC recipients.


\begin{figure}[H]
\DeclareGraphicsExtensions{.pdf,.png,.jpg,.eps}
\centering
\includegraphics[width=2.9in]{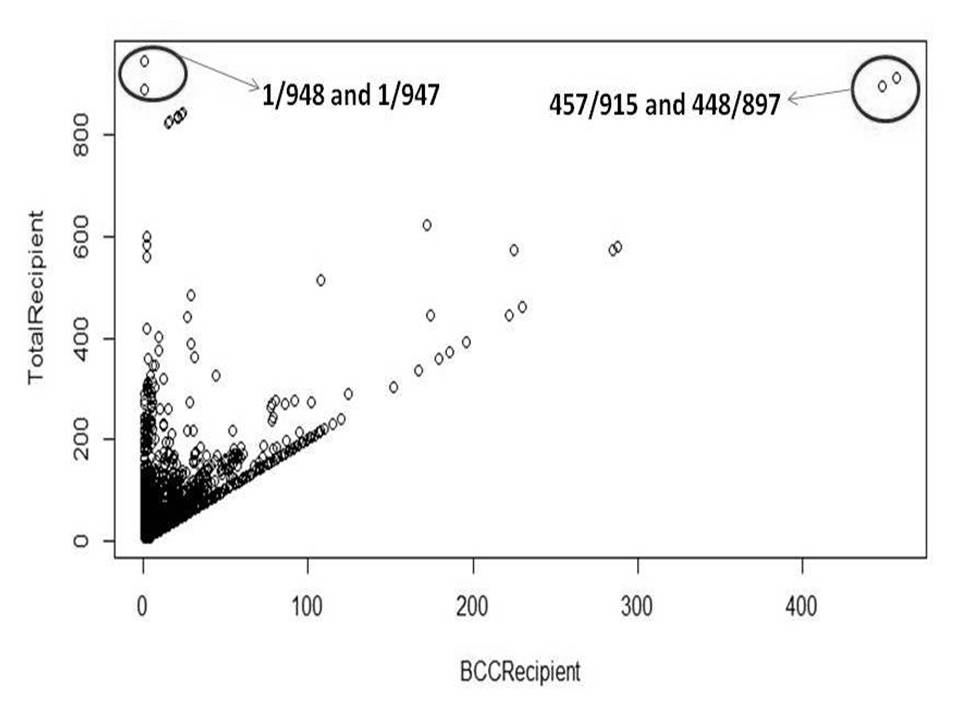}
\captionsetup{singlelinecheck=off,justification=justified}
\caption{The BCC recipients in emails with more than 5 recipients. Note that there are some abnormal scenarios, such as 1 recipient out of 948 being bcc-ed.}
\label{fig:BCCRecipientsinemailswithmorethan5Recipients}
\end{figure}

Based on the statistical findings, the BCC email group was divided into two different groups; emails with more than 5 recipients and emails that had at most 5 recipients.
There were 34195 emails with more than 5 recipients, with a small number of these emails having large bcc recipient lists. The ratios of bcc-ed recipients in the group with more than 5 recipients were calculated. Figure \ref{fig:BCCRecipientsinemailswithmorethan5Recipients} shows that there were some abnormal scenarios detected when BCC recipients in emails with more than 5 recipients were plotted.  For example, 1 recipient out of 948 recipients was bcc-ed. Emails that have large lists of recipients would not typically imply a trust relationship, rather a quirk of the email system or just an information security practice. Thus, this subset was not used for analysis.

Next, the emails that were sent to at most 5 recipients were analysed. There were 26454 emails of this type. On average, out of 5 recipients between 1 and 2 recipients were bcc-ed. Based on this, our analysis was restricted to those emails where only one or two recipients were bcc-ed. We called these the 1-BCC and 2-BCC email networks respectively. The 1-BCC network consists of 5290 nodes 17838 edges whereas the 2-BCC network consists of 3766 nodes and 13486 edges. These two subnetworks are analysed in section \ref{ShortestPathsNetworkAnalysis}. The next section describes the shortest paths network search algorithm.

\vspace*{2mm}

\section{The Shortest Paths Network Search Algorithm}
\label{sec:AnalysisAlgorithm}

The R igraph \cite{Rigraph2006} package was used to create a network graph of all emails with either one or two recipients bcc-ed (the 1-BCC and 2-BCC groups).  The process of obtaining a new network is displayed in Algorithm \ref{alg:MyAlgorithm}.

\subsection{Details of terms used and functions}
\label{sec:TermsandRepresentation}

The following abbreviations of terms are used in the algorithm:

\begin{description}
  \item[$A_C$] \hspace{7pt}  := [Array of criminals]
  \item[$C_i$] \hspace{7pt}  := [$i^{\text{th}}$ Criminal]
  \item[$EC_i$] \hspace{7pt} := [Ego / $i^{\text{th}}$ criminal in the list $A_C$]
  \item[$N_{EC_i}$] \hspace{7pt}  := [Ego / $i^{\text{th}}$ criminal's subnetwork]
  \item[$MI_{N_{EC_i}}$] \hspace{7pt} := [Most Influential Node in $N_{EC_i}$]
  \item[$MM_{N_{EC_i}}$] \hspace{8pt}  := [Middle Man Node in $N_{EC_i}$]
  \item[$OC$] \hspace{7pt}  := [Other Criminals in the ego subnetwork $N_{EC_i}$]
  \item[$R$]  \hspace{7pt} := [Result]
\vspace{1em}
\end{description}

$C_i$ is a criminal and each criminal is stored in an array, $A_C$. For each iteration, we take a criminal $C_i$ from $A_C$ as an ego and refer to it as $EC_i$. An ego can be any suspicious entity. In our BCC email network analysis, $C_i$ refers to the criminals involved in the Enron money laundering crime as stated in Table \ref{table:MoneyLaunderingCriminals}. $N_{EC_i}$ is the $i^{\text{th}}$ criminal's subnetwork, also called the ego subnetwork. An ego subnetwork is a network which comprises of all the vertices reachable from the $i^{\text{th}}$ ego, $EC_i$, all the vertices from which $i^{\text{th}}$ ego, $EC_i$, is reachable and all the links connecting these two sets of vertices. The set of all the vertices reachable from an $i^{\text{th}}$ ego, $EC_i$, is called the out-component and the set of all the vertices from which $i^{\text{th}}$ ego, $EC_i$, is reachable is called the in-component. A vertex $v_{j}$ is reachable from vertex $v_{i}$ if $v_{j}$ was bcc-ed in an email from $v_{i}$. If there exists more than one email from $v_{i}$ to $v_{j}$ where $v_{j}$ was bcc-ed then we simplify it by removing multiple links. $MI_{N_{EC_i}}$ and $MM_{N_{EC_i}}$ are the most influential node and the middle man node in the ego subnetwork $N_{EC_i}$ respectively. $OC$ refers to other criminals in $N_{EC_i}$ not including the ego $EC_i$. $R$ denotes the result that is obtained from each step of Algorithm \ref{alg:MyAlgorithm}. 

Despite the different subject matter exchanged between a sender and a recipient, a directed unweighted BCC shortest paths network graph is formed where the edge between one node to another shows a trust relationship (see Figure \ref{fig:OneBCCSPNConnection}). To start with, we identified the criminals that existed in both the 1-BCC and 2-BCC email groups. In the 1-BCC email network, only Andrew Fastow (686 and 687), Lea Fastow (11010, 11009), Kevin Hannon (10068), Kenneth Rice (9994), Rex Shelby (15224, 15225) and A. Khan (205) exist. Meanwhile, the criminals that exist in the 2-BCC network are Andrew Fastow (686 and 687), Lea Fastow (11010, 11009), Kevin Hannon (10068), Ben Glisan (1369),  Kenneth Rice (9994) and Rex Shelby (15224).  The rest of the criminals, 3 of the 10, didn't appear in either the 1-BCC or the 2-BCC networks.
 
From the same ego subnetwork, $N_{EC_i}$, we also extract the shortest paths from other criminals in the array, $A_C$, to the MI and the MM followed by extracting the shortest paths from the $EC_i$ to the other criminals. In the last step of the algorithm, all the shortest paths are combined to give a shortest path network, showing each criminal's position and the network association pattern. The steps are shown in Algorithm \ref{alg:MyAlgorithm}.


\begin{algorithm}[H]
\caption{Criminal shortest paths network search algorithm}
\label{alg:MyAlgorithm}
{\fontsize{8}{8}\selectfont
A. Store the criminals, $C_i$ to $C_n$  in an array $A_C$ \\
B. Form subnetwork of each ego and follow the steps below until all ego subnetworks have been tested.\\
\begin{minipage}{\linewidth}
\begin{algorithmic}[0]
\vspace*{2mm}
\For{$i=1$ to $n$}
\State 1. Select a criminal, $C_i$ from $A_C$ as ego, $EC_i$.\\
\State 2. Retrieve the ego subnetwork $N_{EC_i}$.\\
\vspace*{4mm}
     \textit{\textbf{(a) connection from ego to MI in ego subnetwork $N_{EC_i}$}}
       \State (i) Find $MI_{N_{EC_i}}$.
          \State (ii) Find the direct path from $EC_i$ to $MI_{N_{EC_i}}$.
             \If{exists}
  			 \State retrieve from graph and output it, then go to (b)...\textbf{R1}
   		   \Else
                 \State go to (a)(iii)
  		   \EndIf
          \\
          \State (iii) Find the shortest path from $EC_i$ to $MI_{N_{EC_i}}$.
           \If{exists}
  			 \State retrieve from graph and output it, then go to (b)...\textbf{R2}
   		\Else
		      \State go to (b)
  		\EndIf
     \\\\
      \textit{\textbf{(b) connection from ego to MM in ego subnetwork $N_{EC_i}$}}
          \State (i) Find $MM_{N_{EC_i}}$.
              \State (ii) Find the direct path from $EC_i$ to $MM_{N_{EC_i}}$.
          		 \If{exists}
  		          \State retrieve from graph and output it, then go to (c)...\textbf{R3}
   			  \Else
                     \State go to (b)(iii)
                  \EndIf
          \\
          \State (iii) Find the shortest path from $EC_i$ to $MM_{N_{EC_i}}$.
           \If{exists}
  	        \State retrieve from graph and output it, then go to (c)...\textbf{R4}
   	     \Else
		   \State go to (c)
  	     \EndIf
	\\\\
        \textit{\textbf{(c) connection from  OC to MI and MM in ego subnetwork $N_{EC_i}$}}
         \For{$i$ $<$ $n$}
           \State (i) Set OC $=$ $\{EC_j  |  j \neq i \}$.
           \State (ii) Find the shortest path from OC to $MI_{N_{EC_i}}$  in $N_{EC_i}$.
           \If{exists}
              \State retrieve from graph and output it, then go to c(ii)...\textbf{R5}
   	     \Else
              \State go to c(iii)
           \EndIf
  	     \\
            \State (iii) Find the shortest path from OC to $MM_{N_{EC_i}}$  in 		
		 $N_{EC_i}$.
	      \If{exists}
  			\State retrieve from graph and output it, then go to (d)...\textbf{R6}
              \Else
   			\State go to (d)
       \EndIf
       \EndFor 
\algstore{myalg}
\end{algorithmic}
\vspace{1em}
\end{minipage}
}
\end{algorithm}

\clearpage

\begin{algorithm}[H]
\ContinuedFloat
\caption{Criminal shortest paths network search algorithm (continued)}
{\fontsize{8}{8}\selectfont
\begin{minipage}{\linewidth}
\begin{algorithmic}[0]
\algrestore{myalg}
\\
 \textit{\textbf{(d) connection from ego to  OC in ego subnetwork $N_{EC_i}$}}
         \For{$i$ $<$ $n$}
          \State (i) Set OC $=$ $\{EC_j  |  j \neq i \}$.
            \State (ii) Find the shortest path from $EC_i$ to OC in $N_{EC_i}$.
              \If{exists}
  		    \State retrieve from graph and output it...\textbf{R7}
               \EndIf
           \EndFor 
\EndFor 
\end{algorithmic}
\vspace{1em}
\end{minipage}
C. Merge \textbf{R1}-\textbf{R7} into a network
}
\end{algorithm}

\vspace*{2mm}

\section{Shortest Paths Networks}
\label{ShortestPathsNetworkAnalysis}

In this section, we present the discovery of the 1-BCC and 2-BCC shortest paths networks and the result of these two networks' analyses. As mentioned before, an edge exists from a node A to a node B only if A sent an email on which B was a BCC recipient. 

\subsection{Discovery of 1-BCC shortest paths trust network}
\label{sec:1-BCC Group}
\vspace*{2mm}
We start by investigating Andrew Fastow's ego subnetwork in the 1-BCC network followed by the other criminals' ego subnetworks. Note that in the 1-BCC network, only one of Andrew Fastow's two ego subnetworks exists, that is andrew.fastow@enron.com (686) exists, but andrew.fastow@ljminvestments.com (687) doesn't. The connection from Andrew Fastow to the MM (16383 - sara.shackleton@enron.com) and the MI (19075 - vince.kaminski@enron.com) were retrieved. This step was repeated for the other criminals in Andrew Fastow's ego subnetwork. Finally the algorithm found the shortest paths from Andrew Fastow to the other 5 criminals in his ego subnetwork. 

\begin{figure}[H]
\DeclareGraphicsExtensions{.pdf,.png,.jpg}
\centering
\includegraphics[width=3.0in]{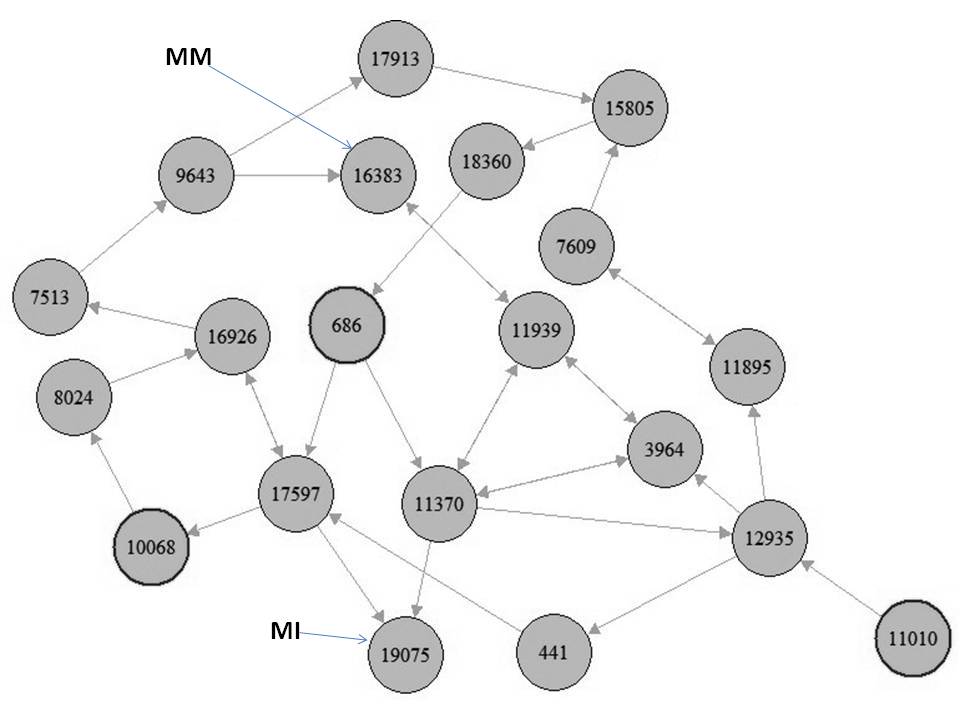}
\captionsetup{singlelinecheck=off,justification=justified}
\caption{The 1-BCC criminals' shortest paths trust network. The MM (16383 - Sara Shackleton) and the MI (19075 - Vince Kaminski) are indicated. All the criminals who occurred in this trust network, Andrew Fastow (686), Lea Fastow (11010) and Kevin Hannon (10068), are highlighted.}
\label{fig:OneBCCSPNConnection}
\end{figure}

Next, two other criminals' ego subnetworks were found; Lea Fastow (11010) and Kevin Hannon (10068). On retrieval of Kenneth Rice (9994) and A.Khan (205)'s ego subnetworks, there were only 2 and 3 nodes respectively. In these small groups, paths don't exist due to two scenarios; betweenness centrality value does not exist or two nodes obtained the same eigenvalue. Similarly, the algorithm drops Lea Fastow (11009) and Rex Shelby (15224, 15225). Following this, the most influential and the middle man were picked from Lea Fastow (11010) and Kevin Hannon (10068)'s subnetworks. In both cases, we obtained the same MI and MM as in Andrew Fastow's ego subnetwork. Combining the results obtained from running algorithm 1, the 1-BCC shortest paths trust network was constructed (see Figure \ref{fig:OneBCCSPNConnection}).

\vspace*{-2mm}

\subsubsection{\textbf{Result (Figure \ref{fig:OneBCCSPNConnection}) of 1-BCC shortest paths trust network}}
\label{1-BCC shortest path network analysis}
The network shows the criminals and the nodes that are closely connected to them. Out of the 6 criminals existing in the 1-BCC network, only Andrew Fastow (686), Lea Fastow (11010) and Kevin Hannon (10068) were extracted. The criminals are located within a range of 2 to 5 hops from the MM and the MI. We also notice that Louise Kitchen (louise.kitchen@enron.com (11370)), Jeff Skilling (jeff.skilling@enron.com (8024)) and Michael (Mike) McConnell (mike.mcconnell@enron.com (12935)) occur the most number of times connecting Andrew Fastow (686), Kevin Hannon (10068) and Lea Fastow (11010) respectively to other nodes. In this shortest paths trust network, all three of these nodes are adjacent to the criminals.

\subsection{Discovery of 2-BCC shortest paths trust network}
\label{sec:2-BCC Group}
\begin{figure}[H]
\centering
\includegraphics[width=2.8in]{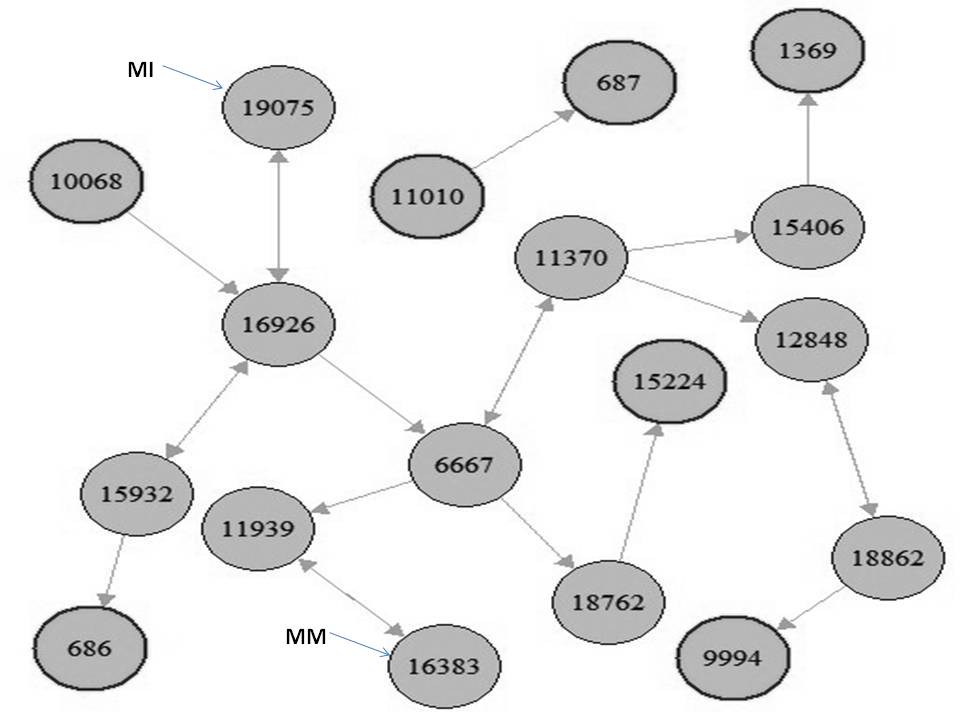}
\captionsetup{singlelinecheck=off,justification=justified}
\caption{The 2-BCC criminals' shortest paths trust network. The MM (16383 - Sara Shackleton) and the MI (19075 - Vince Kaminski) are indicated. All the criminals who occurred in this trust network, Andrew Fastow (686, 687), Lea Fastow (11010), Kevin Hannon (10068), Ben Glisan (1369), Kenneth Rice (9994) and Rex Shelby (15224), are highlighted.}
\label{fig:TwoBCCSPNConnection}
\end{figure}

Next we ran the algorithm on the ego subnetworks of Andrew Fastow and the other criminals in the 2-BCC email group.
Both of Andrew Fastow's (686 and 687) subnetworks were extracted. Here too, the two important central nodes in both of the Andrew Fastow's (686 and 687) subnetworks were (sara.shackleton@enron.com - 16383), the MM and (vince.kaminski@enron.com - 19075), the MI. In both of Andrew Fastow's (686 and 687) ego subnetworks, only Kevin Hannon (10068) was connected to the most influential and middleman. The algorithm was also applied to each of the other 5 criminal subnetworks that exist in the 2-BCC email group. The result of the algorithm is the 2-BCC shortest paths trust network (See Figure \ref{fig:TwoBCCSPNConnection}).\\

\subsubsection{\textbf{Result (Figure \ref{fig:TwoBCCSPNConnection}) of 2-BCC shortest paths trust network}}
\label{2-BCC shortest path network analysis}
All 6 criminals were found within the 2-BCC shortest paths network; they are Andrew Fastow (686, 687), Lea Fastow (11010), Kevin Hannon (10068), Ben Glisan (1369), Kenneth Rice (9994) and Rex Shelby (15224). Out of these, Andrew Fastow (686, 687), Ben Glisan (1369), Rex Shelby (15224) and Kenneth Rice (9994) did not have an out-component but acted as end nodes. The criminals who had out-components were Kevin Hannon and Lea Fastow.  In this trust network, there exists a seperate single connection from Lea Fastow (11010) to Andrew Fastow (687). The only criminal that has a path to the MI and MM is Kevin Hannon in the range of 2 to 4 hops away. The node that connects Kevin Hannon to other nodes the most is Sherri Sera (sherri.sera@enron.com (16926)) followed by Greg Piper (6667).

The position and connections of criminals with other nodes could be used for further investigation. In the next section, we test the ability of the shortest paths network search algorithm on a scenario where an investigator is at the beginning stages of an investigation with no information about who the criminals may be. He only suspects that a money laundering crime is occurring.

\vspace*{10mm}
\section{First Suspect Test}
\label{sec:firstsuspectTest}
The first suspect test replaces all criminals with a group of people who are potentially under suspicion. Each combined shortest paths network in these tests is analysed separately and the number of criminals who occurred are counted. The purpose of these experiments is to find a suitable sparse subgraph to start an investigation.

This money laundering first suspect test is started by investigating the Enron officials who were involved in financial account management. In the financial network, financial managers act as egos. The union of financial managers' network from the BCC network consist of shortest paths from each financial manager (ego) to the MI, to the MM and from one financial manager to another. The financial managers are Sherron Watkins (sherron.watkins@enron.com (16929)), the Head of Enron Global Finance, Andrew Fastow (andrew.fastow@enron.com (686), andrew.fastow@ljminvestments.com (687)), the Enron Chief Financial Officer, Ben Glisan (ben.glisan@enron.com (1369)), the Enron Corporation Treasurer, Rick Causey (rcausey@enron.com (15077)), the Chief Accounting Officer, Jeff McMahon (jeffrey.mcmahon@enron.com (8071)), the Chief Financial Officer of Enron after Andrew Fastow. We notice that this list of financial managers includes two criminals; Andrew Fastow  and  Ben Glisan. We assume that the investigator would not know this in the beginning stage of a crime investigation. The next part shows the result of our first suspect test on the 1-BCC and 2-BCC financial managers' networks.

\subsection{\textbf{Result of first suspect test in 1-BCC financial manager network}}
\label{sec: Firstsuspectin1-BCCFinNetwork}
\begin{figure}[H]
\DeclareGraphicsExtensions{.pdf,.png,.jpg}
\centering
\includegraphics[width=2.5in]{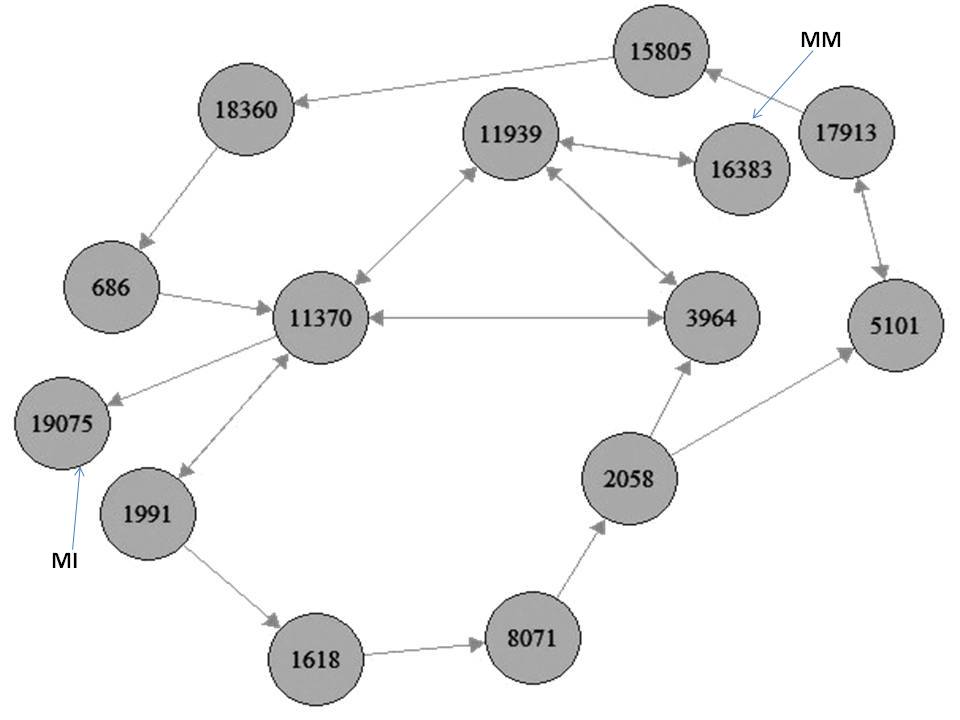}
\captionsetup{singlelinecheck=off,justification=justified}
\caption{The 1-BCC financial managers' shortest paths trust network. The MM (16383 - Sara Shackleton) and the MI (19075 - Vince Kaminski) are indicated. Besides Andrew Fastow (686) who was also a financial manager, no additional criminals occurred in this trust network.}
\label{fig:MergedUnionofFinancial ManagersinOneBCCConnection}
\end{figure}

Figure \ref{fig:MergedUnionofFinancial ManagersinOneBCCConnection} shows the union of financial managers' shortest paths. In this shortest paths trust network two financial managers,  Andrew Fastow (andrew.fastow@enron.com (686)) and Jeff McMahon (jeffrey.mcmahon@enron.com (8071)), occurred. Note that Andrew Fastow was the only criminal identified here with no other criminals appearing. Louise Kitchen (louise.kitchen@enron.com (11370)) was in between Andrew Fastow and other nodes the most number of times. Meanwhile, Bruce Garner (2058) connects Jeff McMahon the most to other nodes. We also compared the nodes that occurred in subgraphs in Figure \ref{fig:MergedUnionofFinancial ManagersinOneBCCConnection} and  in Figure \ref{fig:OneBCCSPNConnection}. The results are discussed in section \ref{sec:DiscussionofResults}.

\vspace{0mm}
\subsection{\textbf{Result of first suspect test in 2-BCC financial manager network}}
\label{sec: Firstsuspectin2-BCCFinNetwork}
\begin{figure}[H]
\DeclareGraphicsExtensions{.pdf,.png,.jpg}
\centering
\includegraphics[width=3.0in]{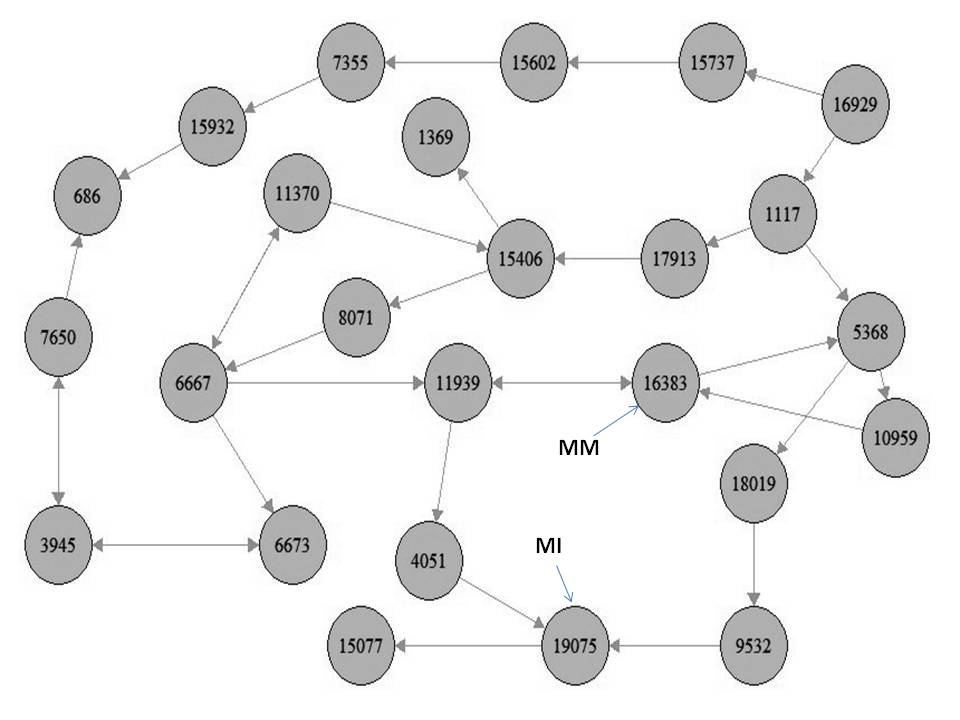}
\captionsetup{singlelinecheck=off,justification=justified}
\caption{The 2-BCC financial managers' shortest paths trust network. The MM (16383 - Sara Shackleton) and the MI (19075 - Vince Kaminski) are indicated. Besides Andrew Fastow (686) and Ben Glisan (1369) who were also financial managers in Enron, no additional criminals occurred in this trust network.}
\label{fig:MergedUnionofFinancial ManagersinTwoBCCConnection}
\end{figure}

\vspace{1mm}
The first suspect test was used on the 2-BCC financial managers' network with the same financial officers used as suspects. Figure \ref{fig:MergedUnionofFinancial ManagersinTwoBCCConnection} depicts the network formed when the union of financial managers' subnetworks is extracted using the shortest paths algorithm. We identified the links from Sherron Watkins (16929) and Jeff McMahon (8071) to other nodes and obtained an unknown email address b..sanders@enron.com (1117) and Greg Piper (6667) as active intermediate nodes respectively. Common nodes are identified from the subgraph in Figure \ref{fig:MergedUnionofFinancial ManagersinTwoBCCConnection} when compared to the subgraph in Figure \ref{fig:TwoBCCSPNConnection}. This is discussed in section \ref{sec:DiscussionofResults}. Again, no additional criminals were found.

\vspace*{1mm}

\section{Discussion of Results}
\label{sec:DiscussionofResults}

\subsection{Common node comparison}
\label{commonnodecomparison}

The subgraph in Figure \ref{fig:MergedUnionofFinancial ManagersinOneBCCConnection} extracted from the 1-BCC email network using the financial managers contains 9 common nodes when compared to the subgraph in Figure \ref{fig:OneBCCSPNConnection}  found using criminals. Some common nodes that we would like to highlight here are Sara Shackleton (16383), Andrew Fastow (686), Louise Kitchen (11370) and Vince Kaminski (19075). Subgraphs that were extracted from the 2-BCC email network show 10 common nodes. Few of them are Ben Glisan (1369), Louise Kitchen (11370), Vince Kaminski (19075), Greg Piper (6667), Andrew Fastow (686), Sara Shackleton (16383). We obtained atleast three criminals in the common nodes' group; 1 criminal (Andrew Fastow (686)) from financial manager’s 1-BCC shortest paths network and 2 criminals (Andrew Fastow (686) and Ben Glisan (1369)) from financial manager’s 2-BCC shortest paths network.

In the first suspect test, starting with the financial managers, no additional known criminals appeared. However, after obtaining these subgraphs, the next step an investigator can attempt is to explore each of these nodes' history. This is possible because these subgraphs are very sparse and small. Even though these people are not yet identified as criminals during this first suspect test, further exploration of these nodes and trust relationships may lead to discovering events that are related to money laundering. 

\subsection{Intermediate node comparison}
\label{Intermediatenodecomparison}

The active intermediate nodes found in section \ref{1-BCC shortest path network analysis} and \ref{2-BCC shortest path network analysis} are Louise Kitchen (11370), Michael (Mike) McConnell (12935), Jeff Skilling (8024), Greg Piper (6667) and Sherri Sera (16926). There were 2 out of 5 persons of interest occurring again when using financial managers as the list of first suspects. These are Louise Kitchen (11370) and Greg Piper (6667). No matter the email contents and number of emails being exchanged, we take all the active intermediaries obtained in section \ref{1-BCC shortest path network analysis}, \ref{2-BCC shortest path network analysis} and section \ref{sec:firstsuspectTest} as persons of interest. We investigated these nodes to see if they had any relationship to important events that occurred during the period leading up to the Enron collapse. 

Jeff Skilling was the president and Chief Operating Officer (COO) of Enron Corporation in December 1996 \cite{anderson2003}. To support Enron's fast growth in the 1990s, Skilling hired the best intellectuals for the company. This accounts for the appointment of Michael (Mike) McConnell \cite{nguyen2013, salter2008innovation} as the Executive Vice President, Technology,  Enron Corp., in July 1999. At the end of 1999, Enron Online came into being and Louise Kitchen, a trader at Enron \cite{nguyen2013, salter2008innovation} was the main person involved in its start-up. McConnell later helped to promote Enron Online \cite{nguyen2013}. The development of the Enron Online was hidden from the COO, Jeff Skilling by Louise Kitchen \cite{sims2003} with the deployment of Enron online being revealed to him only two weeks before it was launched \cite{sims2003}. 

The next person of interest is Greg Piper. Greg Piper (6667) was the Managing Director of Enron NetWorks. He supported the growth of the web based trading introduced by Louise Kitchen \cite{GregPiper}. He was responsible for all of Enron's e-commerce systems development, such as EnronOnline and ClickPaper.com \cite{GregPiper}. Thus, Louise Kitchen and Greg Piper are both connected with Enron Online. Although we can't prove that these identities are suspicious, these interesting and intriguing emails indicate that they should be investigated further. In fact, Louise Kitchen has been identified as an important node in prior research using node neighbourhood search, page rank \cite{yang2010} and rule-based search on the Enron employee's job field \cite{zhao2007}. On the other hand, even though Sheri Sera occurred most frequently between Kevin Hannon and other nodes, history does not indicate that she should be considered suspicious \cite{LoyalAssistantofSkilling}. 

\section{Conclusion and Future Work}
Our shortest paths network search algorithm is able to capture a closely connected trust network. The algorithm managed to show connections between nodes in 2 different scenarios; when an investigator knows all the criminals and when the investigator is at the starting stage and doesn't have any information about the criminals. The analyses conducted in this paper show that when crime is suspected, our algorithm provides a means of identifying possible people to investigate. It is the first step of an investigation: identifying trusted connections between known criminals or financial managers with other active intermediate nodes in a network. Future work includes testing the efficacy of our algorithm on a larger dataset and combining the node ranking with dependency methods to identify the most trusted node of a known source.

\end{document}